\documentclass{llncs}
\usepackage{graphicx}
\usepackage{etoolbox}

\begin{document}

\title{Incidental or influential? - Challenges in automatically detecting citation importance using publication full texts.}
\titlerunning{Incidental or influential?}  
%
\author{David Pride \and Petr Knoth
}
\authorrunning{Pride and Knoth.} 
%
\tocauthor{}
\institute{The Knowledge Media Institute, The Open University, Milton Keynes, UK.\\
\email{\{david.pride, petr.knoth\}@open.ac.uk}
\
\\
}

\maketitle              

\begin{abstract}
This work looks in depth at several studies that have attempted to automate the process of citation importance classification based on the publications' full text. We analyse a range of features that have been previously used in this task. Our experimental results confirm that the number of in-text references are highly predictive of influence. Contrary to the work of Valenzuela et al. (2015) \cite{valenzuela2015identifying}, we find abstract similarity one of the most predictive features. Overall, we show that many of the features previously described in literature are not particularly predictive. Consequently, we discuss challenges and potential improvements in the classification pipeline, provide a critical review of the performance of individual features and address the importance of constructing a large scale gold-standard reference dataset.

\end{abstract}

\section{Introduction}
\vspace*{-\baselineskip}
The three largest citation databases; Google Scholar, Web of Science (WoS) and Scopus all give prominence to citation counts. However, it has been long established that treating all citations with equal weight is counterintuitive. Garfield, the original proponent of the JIF \cite{garfield1972citation}, proposed a range of 15 different reasons a paper may be cited. 

In this paper, we address the problem of identifying influential citations based on publications' full text. The rest of the paper is organised as follows. In Section \ref{rel_work}, we introduce key studies on which our work is based. We then discuss the approach for detecting influential citation, providing a critical analysis of features previously applied in this task in Section \ref{methodology}, selecting a set of three key features for further analysis. We present a comparative study of the identified features in Section \ref{results}, together with the challenges inherent in this task. 

\section{Related Work}
\vspace*{-\baselineskip}
\label{rel_work}
There have been several different methodologies applied to this task, Hou et al. (2011) \cite{hou2011counting} first suggest the idea of using an internal citation count based on the full text of a research paper rather than just the bibliography to determine influence. They demonstrate a positive correlation between the number of times a citation occurs and its overall influence on the citing paper.  Zhu et al. \cite{zhu2015} suggests a range of 40 classification features including both semantic and metric features to determine influence.  Most recently, Valenzuela et al. (2015) \cite{valenzuela2015identifying} made  significant efforts to construct a reference set which was publicly released and which this study relies heavily on. They suggest a range of 12 features, many of which show similarity with those of  \cite{zhu2015}. 

All of the studies under consideration use a range of different features and test them on different datasets. Consequently, getting a deeper understanding of which of the previously suggested features are most effective at this task is needed. 

\section{Methodology}
\label{methodology}
\vspace*{-\baselineskip}
The typical workflow for classifying citation types involves extracting the full text of the manuscript, parsing the text to detect document structure and then applying a classifier trained using machine learning approaches.

In the rest of this section, we describe this workflow concentrating on the selection of features used in the citation type classification task.  

\subsection{Classification features used by prior studies}

One of the overriding aims of this work is to establish which of the previously identified classification features perform most strongly as predictors of citation importance and to use this as a baseline from which to build future work. 

We consider the features presented in the two most recent studies. In \cite{zhu2015} we first see an expansion of the features into a rich range that move beyond simple counting of in-text citations;

We analysed the 40 features used by \cite{zhu2015} and 12 features used in the study of \cite{valenzuela2015identifying}. Of the 40 features suggested by \cite{zhu2015}, a combination of just 4 features resulted in the best performance of the model. Adding features beyond this actually lowered the performance. Out of the 12 features of \cite{valenzuela2015identifying}, we found three features irreproducible (F3, F5\footnote{We attempted to reproduce this feature, but failed due to Valenzuela's dictionary of cue words not being available.}, F12), we were unable to reliably replicate two features due to PDF extraction issues (F2, F6) and we elected not to use two features as they rely on external and potentially changing evidence (F10, F11). Two features we tested (F7, F8) did not produce any significant correlation with the gold standard. 

Of the three remaining features of \cite{valenzuela2015identifying}, we found a complete overlap of two features between \cite{valenzuela2015identifying} and \cite{zhu2015} (F1-countsInPaperWhole, F4-aux\_SelfCite) and a close match on the third (F9-simTitleCore). These three selected features correspond to the best (F1-countsInPaperWhole) feature of Zhu, the worst feature of Valenzuela (F9-simTitleCore) and a third where the opinion regarding the usefulness of this feature was divided between the two studies (F4-aux\_SelfCite). 

\subsection{Classification} 

Using the identified features, we perform a binary incidental / influential classification. WEKA 3 \cite{witten2016data} was selected as the machine learning toolset in our study. 

\section{Results}
\label{results}
\vspace*{-\baselineskip}
\subsection{Dataset}

The dataset released by \cite{valenzuela2015identifying} contains incidental/influential human judgments on  465 citing-cited paper pairs for articles drawn from the 2013 ACL anthology, the full texts of which are publicly available. The judgment for each citation was determined by two expert human annotators and each citation was assigned a label. Using the author’s binary classification, 396 citation pairs were ranked as incidental citations and 69 (14.3\%) were ranked as influential (important) citations.

\subsection{Analysis and comparison of selected features.} 

Our experiments tested a range of features and their efficacy as predictors of citation influence. We achieved the best results using the Random Forests Classifier. We tested the model using bagging with 100 iterations and a base learner, using a 10-fold cross-validation methodology.  The WEKA toolset was used to generate P/R curves for each of the individual features as well as the combination of all the features (Table \ref{all_features}). 

\begin{table}[!h]
\centering
\begin{tabular}{| l | c | c | c | c | c | c |} \hline
Feature                         & P@R=0.05 & P@R=0.1 & P@R=0.3 & P@R=0.5 & P@R=0.7 & P@R=0.9 \\ \hline
F1 & 0.4  & 0.34 & 0.33 & 0.3 & 0.26 &  0.21   \\ \hline
F4 & 0.27 & 0.35 & 0.14 & 0.15 & 0.14 & 0.14   \\ \hline
F9 &  0.46 & 0.49 & 0.21 & 0.2 & 0.18  & 0.16   \\ \hline
All & 0.5 & 0.38 & 0.37 & 0.37 & 0.29  & 0.23    \\ \hline
\end{tabular}
\caption{Interpolated precision at different recall levels for all features for the random forest classifier.}
\label{all_features}
\vspace*{-\baselineskip}
\end{table}

We also measured the correlation between each of the individual features and the classification given by the human annotators. Valenzuela et al. \cite{valenzuela2015identifying} present their results in terms of P/R values for each feature whereas \cite{zhu2015} shows the Pearson correlation with their gold standard. We therefore present the results of our experiments in both formats to allow for accurate comparison. 
\begin{table}[!h]
\centering
\caption{Comparison of results by feature}
\label{Table6}
\begin{tabular}{| l | c | c | c | c |} \hline
& \multicolumn{2}{| c |}{Precision@Recall=0.9} & \multicolumn{2}{|c|}{Pearson $r$} \\ \hline
Feature              & \, Valenzuela et al. \cite{valenzuela2015identifying} \, & \, Our results \, & \, Zhu et al. \cite{zhu2015} \, & \, Our results \, \\ \hline
Direct Citations     & \textbf{0.30} & \textbf{0.21 }&  \textbf{0.330} & 0.281              \\ \hline
Abstract Similarity  & 0.14 & 0.14 &  N/A & \textbf{0.373}       \\ \hline
Author Overlap       & 0.22 & 0.16 &  0.020 & 0.132            \\ \hline
\end{tabular}
\vspace*{-\baselineskip}
\end{table}
Our work confirms the earlier findings reported in \cite{zhu2015} and \cite{valenzuela2015identifying} that the number of direct instances of a citation within a paper is a clear indicator of citation influence. We also find that author overlap, or self-citation, does have value as a classification feature. Contrary to the work of \cite{valenzuela2015identifying} we find that the similarity between abstracts is more predictive of citation influence than previously shown. 

The correlation of this feature with the reference set ($r$=0.373, $p<0.01$, 2-tailed) was the highest of all the features we tested. It is our contention that testing all features using $P/R$ values, at $R0.90$ masks some of the predictive value of those features when the dataset contains only a small number of instances of the influential class. Table 3 shows the precision of the random forests classifier at various recall levels. It can been seen from these results that the classifier initially performs quite well and identifies many of the influential cases, however it has difficulty identifying the last few instances which substantially decreases the classifier's performance at \textit{R}0.90. Using Mean Average Precision (MAP) or a similar metric that provides a single-figure measure of quality across recall levels would be a better choice in this case. 

\subsubsection{Results for Individual Features}
\subsubsection{F1 - Number of Direct Citations :}
\vspace*{-\baselineskip}

This feature is rated as the highest value in terms of predictive ability by \cite{zhu2015} and the second highest by \cite{valenzuela2015identifying}. The latter shows P0.30 at R0.90, however our results demonstrate a slightly lower P value, P0.21 at R0.90. \cite{zhu2015} lists the equivalent 'countsinPaper\_Whole' as the most significant feature of their classifier, with a Pearson correlation coefficient of P0.35. We find a Pearson correlation of P0.28 (significant at the 0.01 level, 2-tailed) for this feature with our dataset. The small difference in this result is likely caused by the differences in the two datasets. Our results therefore confirm that the number of times a citation appears is a strong indicator of that citation's influence. 

\vspace*{-\baselineskip}
\subsubsection{F4 - Author Overlap:}

The results from the two earlier studies for this feature vary considerably. In the results for \cite{valenzuela2015identifying} this is the third ranked 'most significant feature’ with P0.22 for R0.90. We find slightly less precision than \cite{valenzuela2015identifying} for this feature; P0.16 at R0.90. \cite{zhu2015}'s results show little correlation with their gold standard for the similar feature aux\_selfCite (Pearson 0.02). Interestingly, despite the low correlation, this feature was the fourth one selected by their model and did indeed improve the performance of the classifier, albeit only slightly. The experiments with our dataset show a far stronger positive correlation, P0.132 (significant at the 0.01 level, 2-tailed), than that found by \cite{zhu2015}.
\vspace*{-\baselineskip}
\subsubsection{F9 - Abstract Similarity}

Whilst \cite{zhu2015} generated many similarity-based features,  they did not compare citing abstract and cited abstract. This is somewhat surprising as we consider it to be an interesting feature and one that also seems innately logical. The abstract similarity is calculated as the cosine similarity of the tf-idf scores of the two abstracts. By ensuring that the dataset only contains valid data, i.e. the abstract is available for both citing and cited paper, a direct comparison can be made for this feature with \cite{valenzuela2015identifying} who rank this as the lowest of their twelve features, P0.14 at R0.90.

Here our results are the same as \cite{valenzuela2015identifying},with P0.14 at R0.90. However, the Pearson correlation with the gold standard dataset for this feature is the highest of the three features tested in our experiments.  We find a Pearson correlation of 0.373 (significant at the 0.01 level, 2-tailed). This feature was not tested by any of the other earlier studies covered in this work. Our results demonstrate that abstract similarity between citing and cited paper is more predictive of citation influence that previously shown. 
\vspace*{-\baselineskip}
\subsection{The value of complex features.}

Many of the complex features  tested by previous studies have been shown to have little predictive ability in regards to classifying citation function or importance. Some of the most basic features have been shown to offer the strongest potential in identifying important or influential citations. Our research confirms that one of the most simplistic features, i.e. the number of times a citation appears in a paper, is highly predictive of influence. 

Replicating complex features is a non-trivial task unless exact details of how the values for these features were calculated or source code are provided by the original study. We believe that it is essential that the types and values of all features should be provided as part of the research dataset (as opposed to providing just source prior to feature extraction) to serve as a roadmap in replicating them. 

\vspace*{-\baselineskip}
\section{Discussion}
One of the major limitations of this and previous studies is the size of the publicly available, annotated, datasets. The study by \cite{valenzuela2015identifying} uses 465 citing / cited paper pairs. The study by \cite{zhu2015} uses just 100 papers by 40 authors. Due to the unbalanced split between the incidental and influential classes, our complete dataset contained only 61 examples of the positive (influential) class. We argue that due to the relative sparsity of influential citations a much larger reference set is required. This is equally true for negative citations, which have been shown to be even rarer. Training a classifier when the dataset contains so few instances  of the non-neutral classes is problematic and we will address this in future work. The construction of a gold standard dataset containing many thousands of annotated citations, rather than a few hundred, is a significant undertaking but we believe this is a vital step in improving the abilities of the classification models. 

There is a noticeable difference between the datasets used by \cite{zhu2015} and \cite{valenzuela2015identifying} which warrants further study. The \cite{valenzuela2015identifying} dataset annotation was undertaken by two independent annotators and finds significant value in using author overlap as a classification feature. However, the \cite{zhu2015} reference set is annotated by the authors themselves and this study ranks author overlap / self-citation as being of very low importance. It may be that is demonstrates shyness or reticence on behalf of authors to regard their own, earlier, work as being a significant influence. 

Finally we argue that if a citation is considered influential, this original influence remains regardless of external factors or the environment. Therefore, classification features which rely on external and potentially fluid information should be used somewhat cautiously. In future work we will address this issue in greater detail.

\section{Conclusions} 
\vspace*{-\baselineskip}
 Of the features we tested, we find the feature \textit{Abstract Similarity} shows the strongest positive correlation for predicting citation influence. We find \textit{Number of Direct Citations} to also be highly predictive and we find \textit{Author Overlap / Self-Citation} to be less predictive but still valuable as a classification feature. It is important to note that many of the features suggested by earlier studies have been shown to have little predictive ability. 

There is scope for further work surrounding the efficacy and in particular the reproducibility of some of the previously tested classification features. Many of the earlier studies in this domain present results based on sometimes complex and irreproducible features. We contest that this is detrimental to this area of study as a whole and, whilst earlier studies have identified several effective features, having the ability to reproduce them is fundamental to further development in the area of citation classification. 

Whilst it may be a relatively easy task for a human being to identify important or influential citations, building a model to automatically classify these citations with any degree of accuracy is a non-trivial task. A larger scale reference set than those used in this and previous studies is essential, particularly due to the inevitably skewed nature of any dataset of citations annotated according to influence or importance. 
\vspace*{-\baselineskip}
\section{Acknowledgements}
\vspace*{-\baselineskip}
This work has been funded by Jisc and has also received support from the scholarly communications use case of the EU OpenMinTeD project under the H2020-EINFRA-2014-2 call, Project ID: 654021.
\vspace*{-\baselineskip}
\bibliographystyle{splncs}
\bibliography{bibliography.bib}

\end{document}